\documentclass[aps,prl,twocolumn,showpacs,superscriptaddress,amsmath]{revtex4}

\usepackage{graphicx}

\begin{document}


\title{Model for Non-Gaussian Intraday Stock Returns}



\author{Austin Gerig}
\email{gerig@santafe.edu}
\affiliation{School of Finance and Economics, University of Technology, Sydney, Broadway, NSW 2007, Australia}
\affiliation{Santa Fe Institute, 1399 Hyde Park Road, Santa Fe, NM 87501}

\author{Javier Vicente}
\affiliation{Santa Fe Institute, 1399 Hyde Park Road, Santa Fe, NM 87501}

\author{Miguel A. Fuentes}
\affiliation{Santa Fe Institute, 1399 Hyde Park Road, Santa Fe, NM 87501}
\affiliation{Center for Advanced Studies in Ecology and Biodiversity, Facultad de Ciencias Biol\'{o}gicas, Pontificia Universidad Cat\'{o}lica de Chile, Casilla 114-D, Santiago CP 6513677, Chile}
\affiliation{Statistical and Interdisciplinary Physics Group, Centro At\'omico Bariloche, Instituto Balseiro and CONICET, 8400 Bariloche, Argentina}


\date{\today}

\begin{abstract}

Stock prices are known to exhibit non-Gaussian dynamics, and there is much interest in understanding the origin of this behavior.  Here, we present a model that explains the shape and scaling of the distribution of intraday stock price fluctuations (called intraday \emph{returns}) and verify the model using a large database for several stocks traded on the London Stock Exchange.  We provide evidence that the return distribution for these stocks is non-Gaussian and similar in shape, and that the distribution appears stable over intraday time scales.  We explain these results by assuming the volatility of returns is constant intraday, but varies over longer periods such that its inverse square follows a gamma distribution.  This produces returns that are Student distributed for intraday time scales.  The predicted results show excellent agreement with the data for all stocks in our study and over all regions of the return distribution.

\end{abstract}

\pacs{89.65.Gh, 05.45.Tp}


\maketitle


It is well-known that the probability distribution of stock returns is non-Gaussian\cite{Mandelbrot63, Fama65}.  The distribution is fat tailed, which means that extreme price movements occur much more often than predicted given a Gaussian model.  There is considerable interest in determining the origin of non-Gaussian returns, and a large number of recent papers on the subject have been written by physicists\cite{Mantegna95, Gopikrishnan99, Plerou99, BouchaudPotters03, Plerou00, Ramos01, Ausloos03, Tsallis03, Gabaix03, Kozuki03, Farmer04, Silva04, Duarte05, Weber06, Gillemot06, Weber07, Silva07, Bassler07, Gu08}.  These studies often attempt to fit the shape of the return distribution and to determine how it scales in time.  The shape and scaling of the distribution are important because they provide information about the underlying process that is driving asset prices.  In addition, understanding the true distribution of returns is important for asset allocation, risk management, and option pricing.

In this paper, we present evidence that the return distribution for stocks is non-Gaussian, similar across stocks, and stable in shape for intraday time scales.  We show that these results are due to specific properties of the scale of individual returns (called \emph{volatility}), and that the similarity of these properties across stocks allows for their return distributions to collapse onto one curve.  This work is related to the large literature on stochastic volatility models\cite{Shephard05}, and specifically to one of the original papers that suggested such a model to explain the non-Gaussian behavior of returns\cite{Praetz72}.  In that paper, the return distribution was assumed to be a mixture of Gaussian distributions with variances that are inverse gamma distributed -- this produces returns that are Student distributed.  Here, we extend this result by assuming that volatility is slowly varying.  This produces returns that are Student distributed throughout intraday time scales. 

Two explanations for the non-Gaussian shape of the return distribution are often mentioned in the literature.  Our model is an example of the \emph{mixture-of-distributions} hypothesis, which states that return distributions are a mixture of Gaussian distributions with different variances\cite{Fama65, Press67, Praetz72, Clark73, Blattberg74}.  Several papers have suggested different explanations for why the variance changes.  The most popular explanation is that fluctuations in the rate of trade underlie these changes\cite{Mandelbrot67, Press67, Clark73, Ane00, Silva07}.  Here, we measure time in increments of events rather than in clock increments and show that the return distribution exhibits interesting properties without considering the rate of trading.  This is supported by previous work that reports fluctuations in the size of returns dominate those in trading\cite{Plerou00, Farmer04, Weber06, Gillemot06}.

The other explanation for the non-Gaussian shape of the return distribution is known as the \emph{stable Paretian} hypothesis -- this states that returns are pulled independently and identically from a stable or truncated stable distribution\cite{Mandelbrot63, Mantegna95}.  Although a non-Gaussian stable distribution can also be described as a mixture of Gaussian distributions with different variances\cite{Blattberg74}, the stable Paretian hypothesis is considered a separate hypothesis because it explains how the return distribution can retain its non-Gaussian shape for long time intervals without violating the assumption of independent and identically distributed (IID) returns.  Here, we show that the apparent stability of the non-Gaussian shape is not due to a stable distribution but instead is due to a slowly fluctuating volatility, which violates the IID assumption.  This is supported by previous work that reports shuffling volatility removes the fat tails of the return distribution for longer time scales\cite{Gopikrishnan99, Viswanathan03}.

To begin our analysis, we define the $t^{th}$ return as the difference in logarithmic price from time $t$ to time $t+\tau$,
\begin{equation}
r_t(\tau) = \ln{(p_{t+\tau})} - \ln{(p_t)},
\end{equation}
where the price $p_t$ is the midpoint price between the best bid price and offer price in the market (these prices are known as \emph{quotes}).  There are several ways to set the unit of the time index, $t$.  Here, we study returns over the finest possible time scale, which we call \emph{midpoint time}.  In midpoint time, $t$ is updated whenever an event causes a change in the midpoint price.

We model individual returns, $r_t(\tau=1)$, as a discrete time stochastic process with a fluctuating variance, 
\begin{equation}
r_t = \sigma_t \xi_t,
\label{eq.arch}
\end{equation}
where $\xi_t$ is an IID Gaussian $N(0,1)$ random variable and $\sigma_t^2$ is the local variance of the process ($\sigma_t$ is the standard deviation of returns at time $t$ and is commonly called \emph{volatility}).  We neglect any drift for returns, which is small on the time scales we study here.  In the econometrics literature, Eq.~\ref{eq.arch} is the standard form for an autoregressive conditional heteroskedasticity (ARCH) model\cite{Bollerslev94}.  Such models can be interpreted as a diffusion process with a time dependent diffusion parameter (in our case $D=\sigma_t^2$)\cite{Plerou00, Weber06}.

As originally noted by Mandelbrot\cite{Mandelbrot63}, $\sigma_t$ slowly varies in financial markets.  This property is now commonly called \emph{clustered volatility}\cite{Bollerslev94}, and its cause remains unknown.  In our model, we assume that $\sigma_t$ is sufficiently slow varying, such that we can treat it as a constant over intraday time scales.  The consequences of this assumption are discussed and empirically validated later.  Replacing $\sigma_t$ with its local constant value $\sigma$, individual returns can be approximated, $r_t \approx \sigma \xi_t$.  We define the variable $\beta$ as the inverse squared volatility, $\beta\equiv1/\sigma^2$, so that the return distribution can be written,
\begin{equation}
P(r,\tau|\beta) = \sqrt{\frac{\beta}{2\pi \tau}}\exp{\left(-\frac{\beta r^2}{2 \tau}\right)}.
\label{eq.daydist}
\end{equation}
Therefore, within our model, the distribution of returns on any single day is a Gaussian with variance $1/\beta=\sigma^2$.  Because $\beta$ can vary at longer time scales, the return distribution observed with data pulled from many different days is obtained by marginalizing over $\beta$,
\begin{equation}
\mathcal{P}(r,\tau) = \int{P(r,\tau|\beta) f(\beta) d\beta},
\label{eq.return_dist}
\end{equation}
where $f(\beta)$ is the distribution of $\beta$.  This \emph{mixture-of-distributions} formulation was originally suggested several decades ago to explain the non-Gaussian shape of the return distribution.  As presented here, it is similar to the recent field of superstatistics, where the statistics of physical systems are separated by different time scales and stationary distributions are derived from the superposition of these statistics\cite{Beck01, Beck03, Kozuki03}.

Motivated by the empirical data below, we assume that $\beta$ is gamma distributed, $f(\beta|a,b) = \frac{b^a}{\Gamma[a]}\beta^{a-1}e^{-b\beta}$.  This is consistent with previous empirical work that reports that $\sigma_t$ is inverse-gamma distributed\cite{Micciche02}.  There are several economic explanations for why the inverse variance might have this distribution\cite{PlatenHeath06, Borland05}, which we discuss later.  In more general terms, a gamma distribution is one of several distributions with positive support that can be derived from universality arguments\cite{Beck05}.  Carrying out the marginalization above gives the following for the distribution of returns,
\begin{equation}
\mathcal{P}(r,\tau) = \frac{\Gamma[a+(1/2)]}{\Gamma[a]}\frac{1}{\sqrt{2\pi b \tau}}\left(1+\frac{r^2}{2b\tau}\right)^{-[a+(1/2)]},
\label{eq.student}
\end{equation}
which is a variant of the Student's $t$-distribution.  Note that the return distribution is determined solely by the two parameters ($a$ and $b$) from the distribution of the inverse squared volatility, $\beta$, and that it remains a Student's $t$-distribution for different $\tau$.  The distribution appears stable, despite being outside the stable regime, because volatility is assumed constant for these time scales.

To facilitate the presentation of the empirical results below, we define the following normalized variables:
\begin{eqnarray}
\xi^* & = & r_t(\tau)/\sqrt{\tau/\beta}, \\
r^* & = & r_t(\tau)/\sqrt{b\tau}, \\
\mathcal{P}^* & = &  (\Lambda\mathcal{P})^{1/[a+(1/2)]},
\end{eqnarray}
where $\Lambda=\sqrt{2\pi}\Gamma[a]/\Gamma[a+(1/2)]$.  These normalizations allow results for different time scales and different stocks to collapse on a single curve.

\begin{figure*}[htb]
\centering
\includegraphics[height=2.4in]{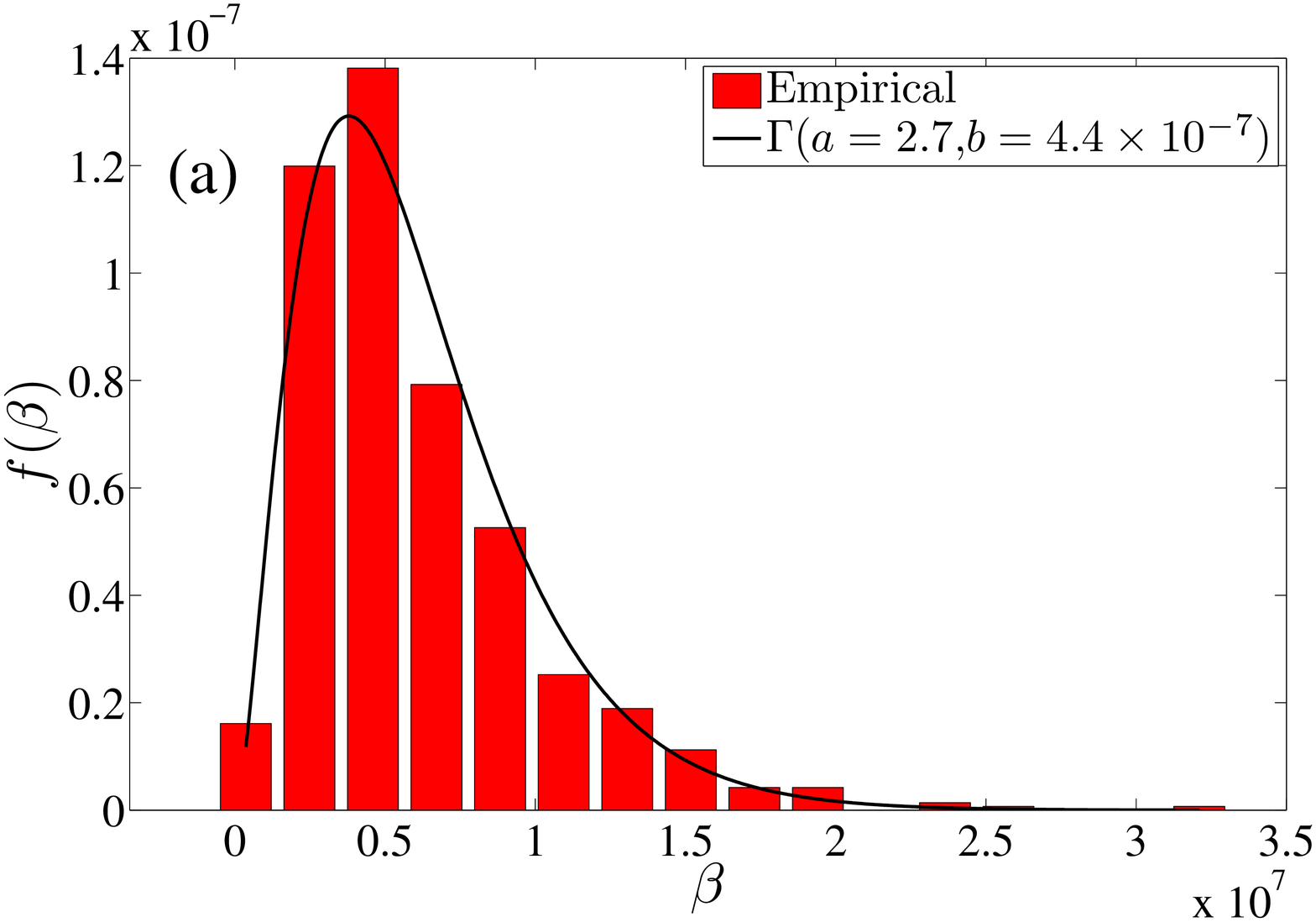}
\includegraphics[height=2.4in]{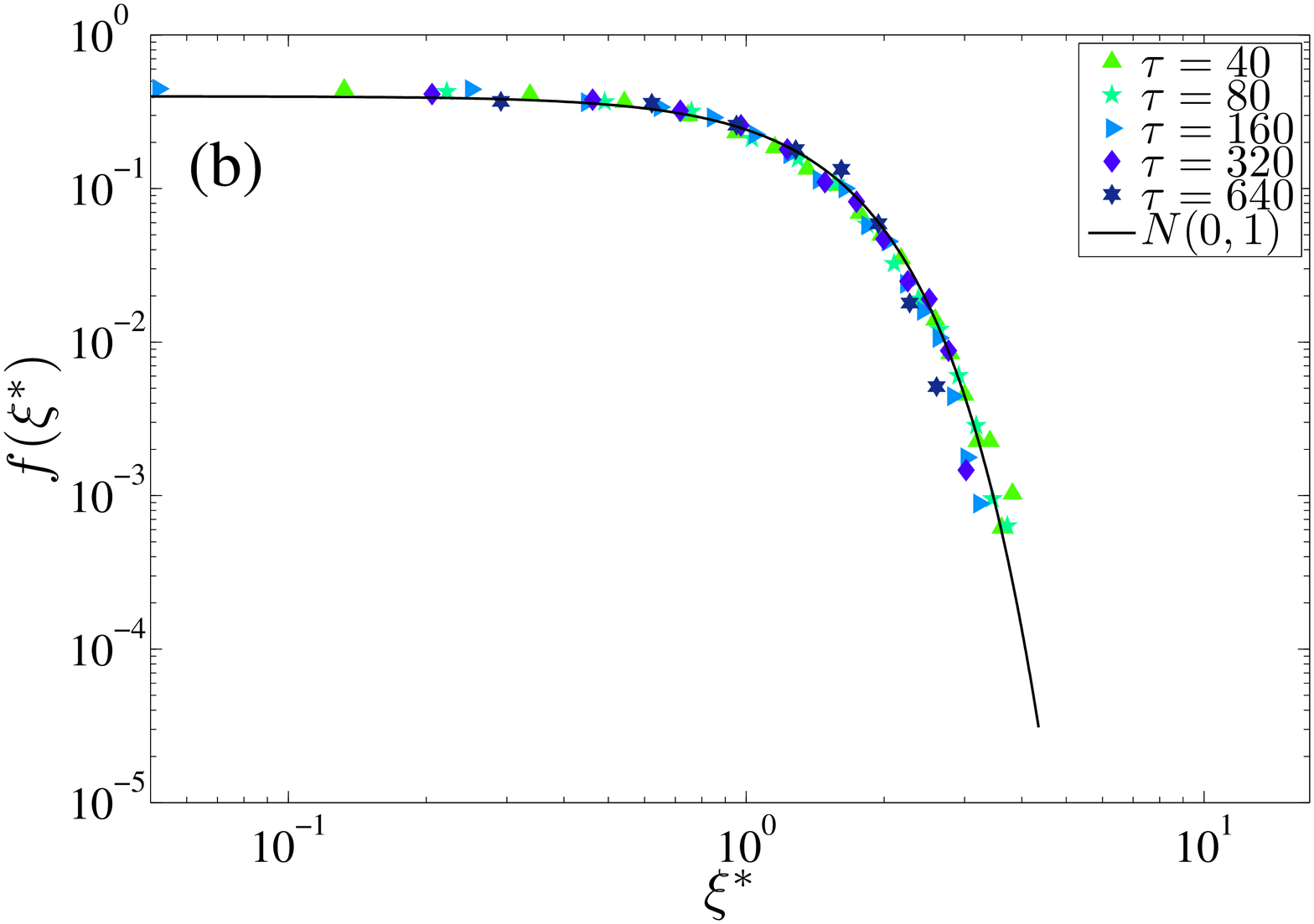}
\includegraphics[height=2.4in]{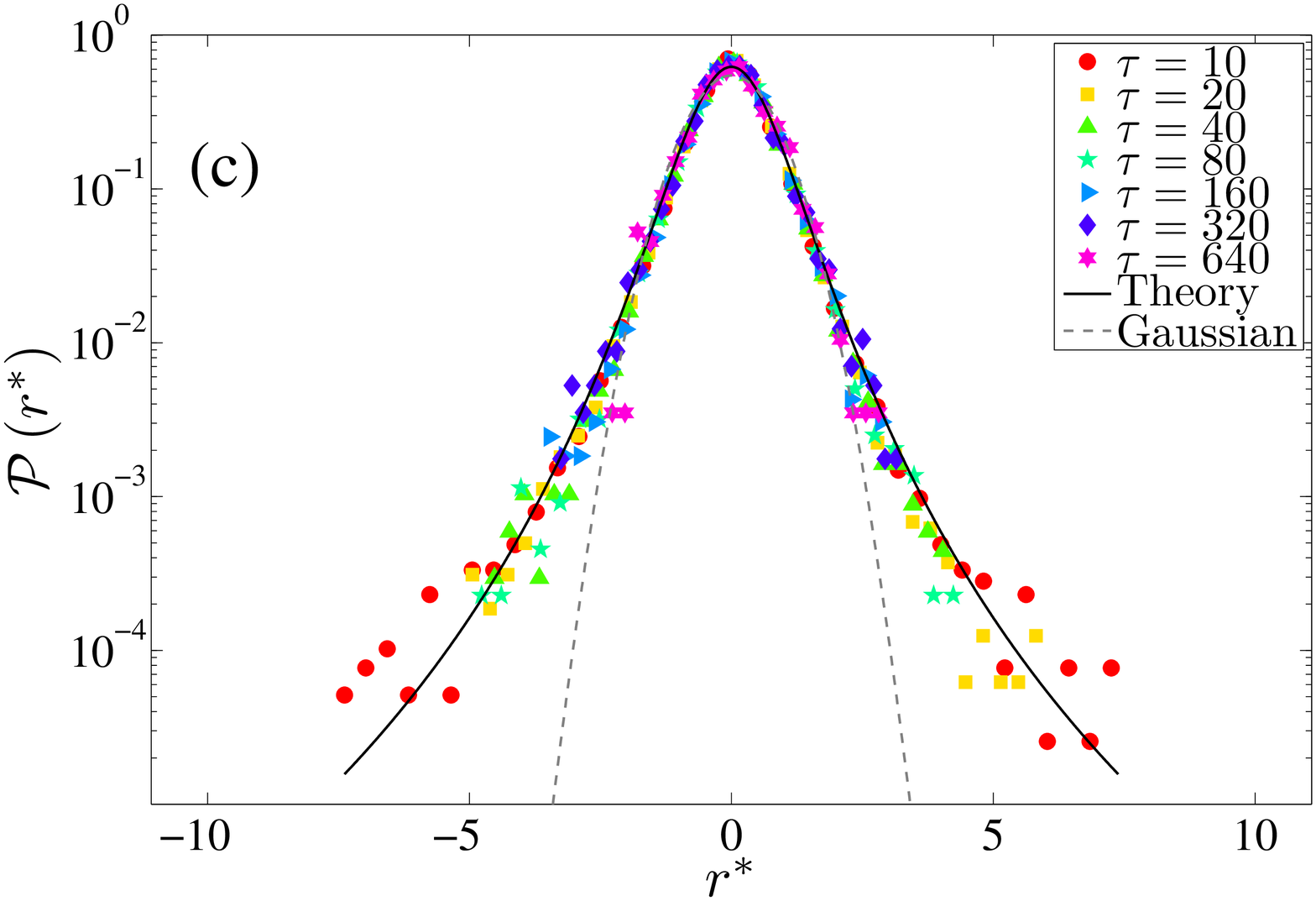}
\includegraphics[height=2.4in]{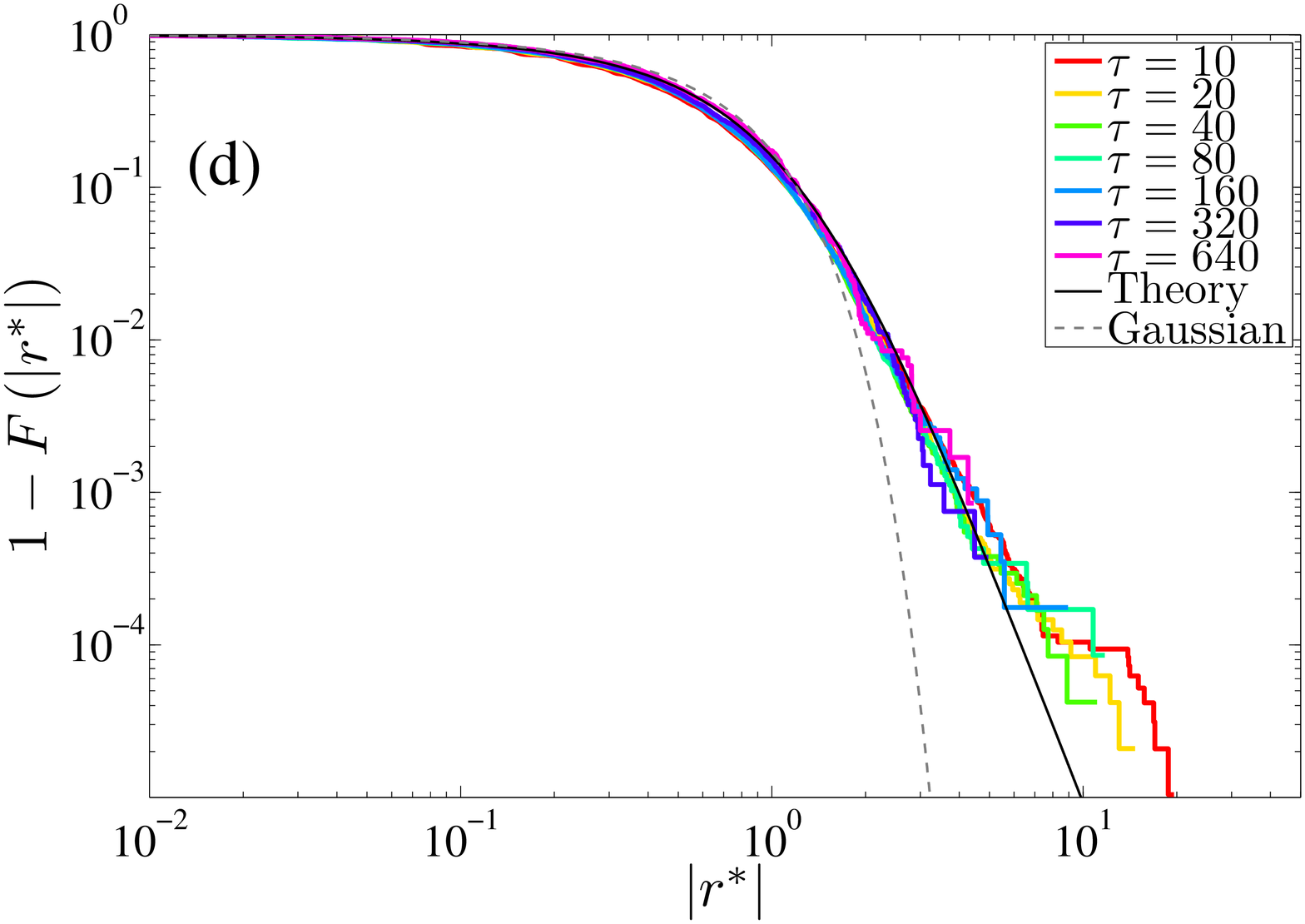}
\caption{Several plots for the stock AstraZeneca (AZN).  (a) The probability density of daily $\beta$ fit by a gamma distribution.  (b) The probability density of $\xi^*$ for different $\tau$ compared to $N(0,1)$.  (c) The probability density of returns for different $\tau$ compared to theory.  (d) The cumulative distribution of returns for different $\tau$ compared to theory.}
\label{fig.AZN_gamma_student}
\end{figure*}

To test the above model, we present empirical results for 5 stocks traded on the London Stock Exchange (LSE) from the period May 2, 2000 to December 31, 2002.  There are 675 trading days during this period.  The stocks are AstraZeneca (AZN), Lloyds TSB (LLOY), Prudential Plc (PRU), Reuters (RTR), and Vodafone Group (VOD).   Our dataset contains information about the complete on-book market -- including all on-book transactions, order placements, and cancellations\footnote{The LSE also contains an off-book market where large transactions take place between nonanonymous counterparties.  We do not use data from the off-book market because the time stamps are unreliable and because public quotes do not exist in this market.}.  We truncate the first 30 minutes of market activity to remove the effects of price discovery at the beginning of the day.  In all of the results for aggregate numbers of events, we choose nonoverlapping intervals.  We only consider intraday returns and do not include returns measured across days.

\begin{figure*}[htb]
\centering
\includegraphics[height=2.4in]{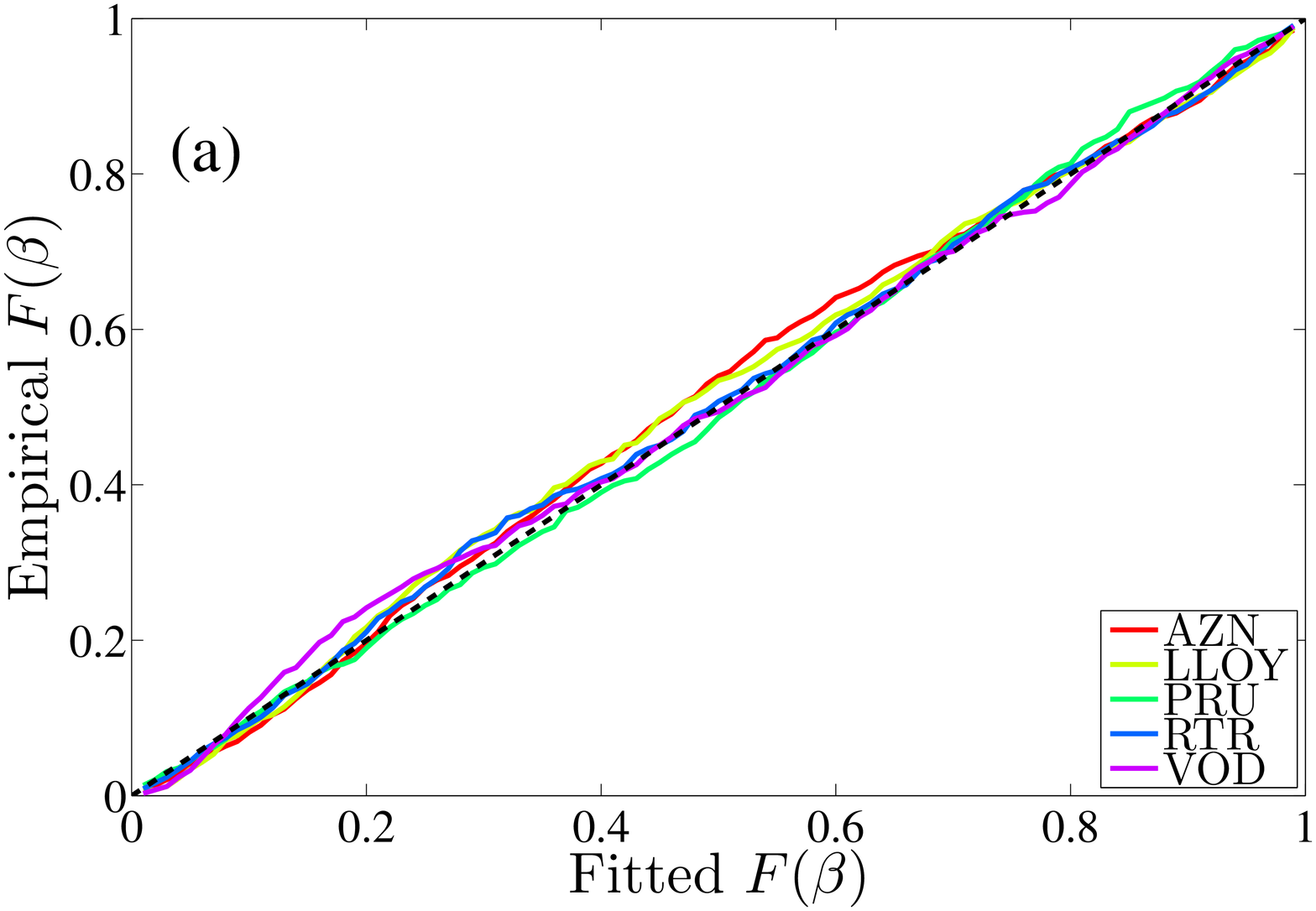}
\includegraphics[height=2.4in]{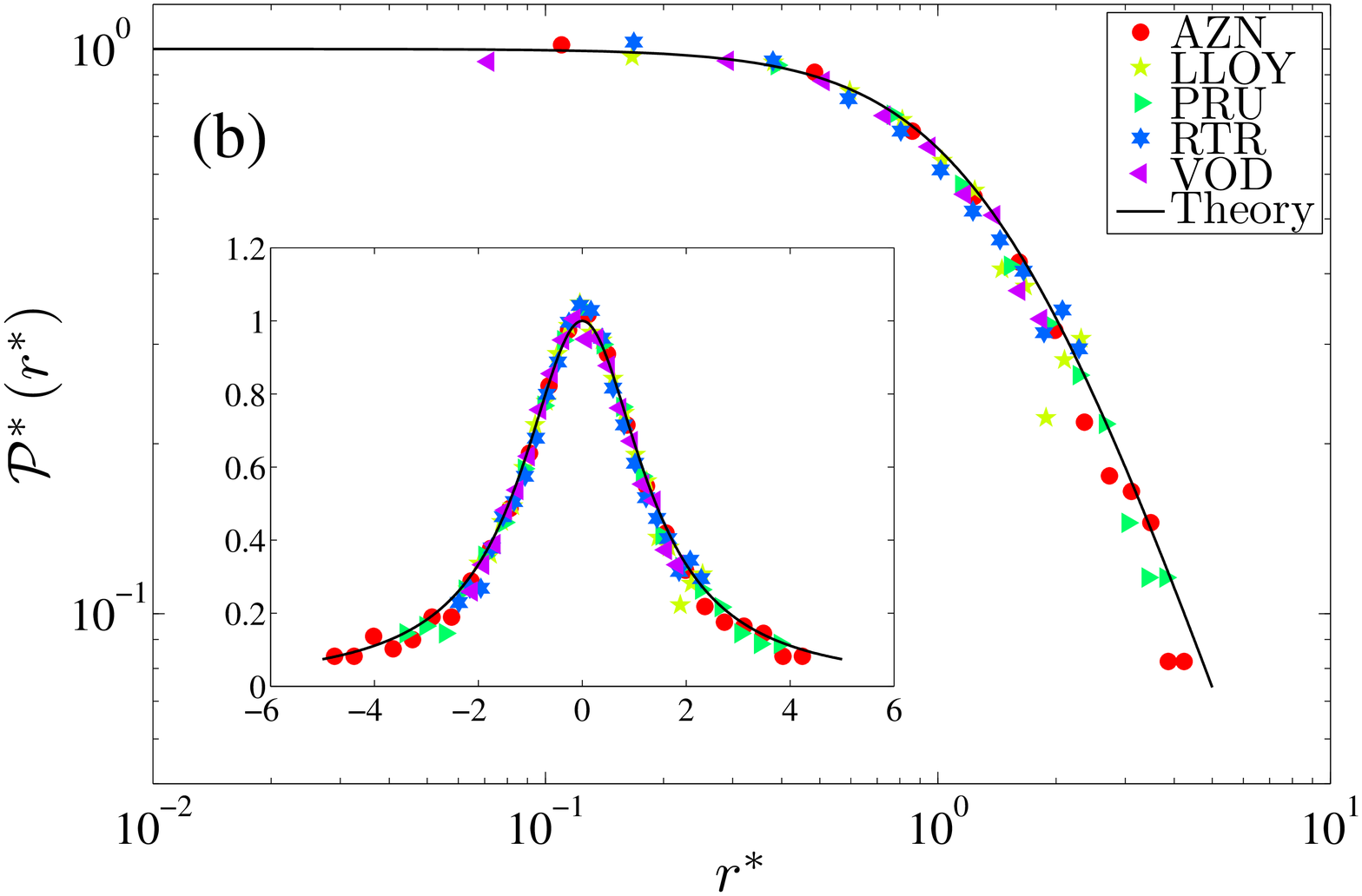}
\caption{(a) The cumulative distribution of $\beta$ compared to the cumulative distribution from the best fit to a gamma distribution for all the stocks in our study.  (b) The normalized probability density of returns (with $\tau=80$) compared to theory for all the stocks in our study.}
\label{fig.ALL_collapse}
\end{figure*}

In the plots below, we compare empirical results to those predicted by the above model.  Others have successfully fit returns to a Student's $t$-distribution (called a $q$-Gaussian or Tsallis distribution in some papers)\cite{Praetz72, Blattberg74, Hurst97, Ramos01, Ausloos03, BouchaudPotters03, Tsallis03, Duarte05, Gu08}.  Note that we do not fit the return distribution here, but instead determine the two parameters, $a$ and $b$, from a maximum likelihood estimate given the daily $\beta$'s \footnote{To obtain a better statistic of the daily $\beta$, we include in our estimate the scaled measurement of volatility for all time scales within one day.}.  These parameters then set the return distribution for intraday time scales.

In Fig.~\ref{fig.AZN_gamma_student} we present results for the stock AZN; although not shown, the results for the other stocks in our study are similar in appearance.  In Fig.~\ref{fig.AZN_gamma_student}(a), we plot the probability density function of $\beta$.  We overlay the plot with the best-fit gamma distribution and we report the parameters for this fit in the figure legend and also in Table~\ref{table.one}.  In Fig.~\ref{fig.AZN_gamma_student}(b), we show the probability density of $\xi^*$ for $\tau=40$ to $\tau=640$ in log-log coordinates.  This is compared to a normal distribution with zero mean and unit variance -- which is assumed in our model.  At time scales shorter than $\tau=40$, which are not shown, the distribution of $\xi^*$ is leptokurtic but with finite variance.  As seen in the figure, the distribution approaches a Gaussian for time scales, $\tau>40$.  That $\xi^*$ is Gaussian distributed was also reported for daily time scales in\cite{Andersen01}.  In Fig.\ref{fig.AZN_gamma_student}(c) we plot the scaled return probability density for $\tau=10$ to $\tau=640$ in semi-log coordinates.  Using the parameters $a$ and $b$, we predict the full probability distribution of returns as derived in Eq.~\ref{eq.student} and overlay this prediction on the plot.  We focus on the tails of the distribution in Fig.\ref{fig.AZN_gamma_student}(d) by plotting the scaled cumulative distribution function for the unsigned returns $F(|r^*|)$ in log-log coordinates.  As seen in both plots, the distributions collapse both in the central region and in the tails and are well described by the predicted curve.  In our model, the collapse occurs because volatility is assumed constant intraday.  For comparison purposes we fit a Gaussian distribution to the data for $\tau=80$ and plot this in Figs.\ref{fig.AZN_gamma_student}(c,d).

\begin{table}[htb]
\centering
\begin{tabular}{lcccccccccc}
\hline
\hline
Security	&&Events		&&Events/Min 	&&$a$		&&$b$ \\
\hline
AZN		&&962516		&&3.0 	&&2.7		&&$.44\times 10^{-6}$\\
LLOY	&&746845		&&2.3		&&3.4		&&$1.1\times 10^{-6}$\\
PRU		&&583792		&&1.8		&&2.6		&&$1.5\times 10^{-6}$\\
RTR		&&653915		&&2.0		&&3.9		&&$3.6\times 10^{-6}$\\
VOD		&&770352		&&2.4		&&3.9		&&$2.1\times 10^{-6}$\\
\hline
\hline
\end{tabular}
\caption{Table of parameters for the five stocks studied.}
\label{table.one}
\end{table}

In Fig.~\ref{fig.ALL_collapse}(a), we plot the empirical cumulative distribution of $\beta$ versus the fitted cumulative distribution for all 5 stocks.  This plot is created by first fixing the value of the fitted $F(\cdot)$, calculating $\beta$ at this point, and then plotting the value of the empirical $F(\cdot)$ for this $\beta$.  The plot is similar to a Q-Q plot -- when the empirical distribution follows the fitted distribution exactly, the curve will lie on the $45^{\circ}$ line.  In Fig.~\ref{fig.ALL_collapse}(b), we plot the normalized probability density $\mathcal{P}^*(r^*)$ with $\tau=80$ for the five stocks in our study.  The data from all five stocks collapse on the curve.

Taken together, these results suggest that slow, significant fluctuations in volatility produce the interesting features of the intraday return distribution.  In this paper, we do not provide an explanation for why volatility has the properties that we have assumed.  We note, however, that there exists a general class of stochastic volatility models that produce volatilities that are inverse gamma distributed\cite{PlatenHeath06}.  The generalized ARCH (GARCH)(1,1)\cite{Nelson90} model and the \emph{3/2 model} are two specific examples.  Variations of the GARCH(1,1) model can be motivated by simple feedback mechanisms for volatility and have been shown to produce similar results to our empirical findings\cite{BouchaudPotters03, Borland05}.  The \emph{3/2 model} is known to be a by-product of a one-dimensional diffusion equation for prices: a squared Bessel process of dimension four\cite{PlatenHeath06}.  This process describes the dynamics of a growth optimal portfolio with deterministic drift and can be motivated by straightforward economic arguments\cite{Platen01, PlatenHeath06}.  

In this paper, we have presented a model for individual stock returns that reproduces the shape and scaling of the intraday return distribution for a collection of stocks from the London Stock Exchange.  Our model decomposes individual returns into the product of two terms: a Gaussian term and a slowly varying volatility term.  On any single day, volatility is relatively constant so that returns are well described by Gaussian fluctuations.  Across many days, however, fluctuations in the volatility term dominate and lead to a non-Gaussian distribution for returns.  The resulting distribution -- a Student's $t$-distribution -- appears stable for short to intermediate time scales despite being outside the stable regime.  This occurs because volatility is slowly varying and therefore not IID.  Thus, we can explain both the shape and apparent stability of the distribution, two results that previously seemed to contradict one another and that individually could be used to support one or the other competing hypotheses for non-Gaussian returns.  Finally, we find that the statistical properties of volatility for different stocks are similar, allowing for a single representation of the return distribution for intraday time scales.

\begin{acknowledgments}
This work was supported by the US National Science Foundation Grant HSD-0624351.
\end{acknowledgments}

\end{document}